\documentclass[aps,prb,showpacs,twocolumn,showpacs]{revtex4}

\usepackage{graphicx}
\usepackage{amsmath}
\usepackage{amssymb}
\usepackage{color}

\definecolor{lila}{rgb}{0.5,0,1}
\newcommand{\bnen}{\begin{equation}}
\newcommand{\eden}{\end{equation}}
\newcommand{\bean}{\begin{eqnarray}}
\newcommand{\eean}{\end{eqnarray}}

\newcommand{\bna}{\begin{array}}
\newcommand{\eda}{\end{array}}

\begin{document}

\title{Periodic Anderson model with correlated conduction electrons: \\
variational and exact diagonalization study}

\author{I. Hagym\'asi$^{1,2}$}
\author{K. Itai$^1$}
\author{J. S\'olyom$^{1}$}

\affiliation{$^1$Research Institute for Solid State Physics and Optics of the 
Hungarian Academy of Sciences, Budapest, H-1525 P.O. Box 49, Hungary\\
$^2$Institute of Physics, E\"otv\"os University, Budapest, P\'azm\'any P\'eter s\'et\'any 1/A, 
H-1117, Hungary}

\date{\today}

\begin{abstract}
We investigate an extended version of the periodic Anderson model (the so-called
periodic Anderson-Hubbard model) with the aim to understand the role of interaction 
between conduction electrons in the formation of the heavy-fermion and mixed-valence 
states. Two methods are used: (i) variational calculation with the Gutzwiller wave 
function optimizing numerically the ground-state energy and (ii) exact diagonalization
of the Hamiltonian for short chains. The $f$-level occupancy and the 
renormalization factor of the quasiparticles are calculated as a function of the 
energy of the $f$-orbital for a wide range of the interaction parameters. The results 
obtained by the two methods are in reasonably good agreement for the periodic 
Anderson model. The agreement is maintained even when the interaction between 
band electrons, $U_d$, is taken into account, except for the half-filled case. 
This discrepancy can be explained by the difference between the physics of the one-
and higher dimensional models. We find that this interaction shifts and widens 
the energy range of the bare $f$-level, where heavy-fermion behavior can be 
observed. For large enough $U_d$ this range may lie even above the bare 
conduction band. The Gutzwiller method indicates a robust transition from 
Kondo insulator to Mott insulator in the half-filled model, while $U_d$ 
enhances the quasi-particle mass when the filling is close to half filling.    

\end{abstract}


\maketitle

\section{Introduction}

Rare-earth materials exhibit numerous remarkable phe\-nom\-e\-na such as 
heavy-fermion behavior, valence fluctuations, and unconventional superconductivity. 
The simplest model that can account for these phenomena is the periodic Anderson 
model (PAM), where mobile conduction electrons in a broad band of width $W$ can 
hybridize with immobile $f$-electrons sitting at the lattice sites. The Coulomb
repulsion is taken into account between the $f$-electrons only. Written in a mixed, 
Bloch and Wannier representation, this model is defined by the Hamiltonian
\begin{equation}   \begin{split}
     \mathcal{H} = & \sum_{\boldsymbol{k},\sigma}\varepsilon_d(\boldsymbol{k})
       \hat{d}_{\boldsymbol{k}\sigma}^{\dagger}
	   \hat{d}^{\phantom \dagger}_{\boldsymbol{k}\sigma}
	   +\varepsilon_f\sum_{j,\sigma}\hat{n}^f_{j\sigma} \\
	   & - V\sum_{j,\sigma}(\hat{f}_{j\sigma}^{\dagger}
	  \hat{d}^{\phantom \dagger}_{j\sigma}
    +\hat{d}_{j\sigma}^{\dagger} \hat{f}^{\phantom \dagger}_{j\sigma})    
     +U_f\sum_{j}\hat{n}^f_{j\uparrow} \hat{n}^f_{j\downarrow} , 
\label{PAM:Hamiltonian}
\end{split}
\end{equation}
where $\hat{d}_{\boldsymbol{k}\sigma}^{\dagger}$ 
($\hat{d}_{\boldsymbol{k}\sigma}^{\phantom \dagger}$) 
is the creation (annihilation) operator of conduction electrons with wave 
vector $\boldsymbol{k}$ and spin $\sigma$, while $\hat{f}_{j\sigma}^{\dagger}$ 
($\hat{f}_{j\sigma}^{\phantom\dagger}$) denotes the creation 
(annihilation) operator of $f$-electrons at site $\boldsymbol{r}_j$ in an 
arbitrary dimensional lattice with $N$ lattice sites,
$\hat{n}^f_{j\sigma}=\hat{f}_{j\sigma}^{\dagger}
  \hat{f}_{j\sigma}^{\phantom\dagger}$ is the number operator of $f$-electrons
at site $\boldsymbol{r}_j$, and $\hat{n}^d_{j\sigma}$ is defined similarly. The 
hybridization matrix element between $f$- and $d$-states is denoted by $V$, and 
$U_f$ is the strength of the on-site Coulomb repulsion between $f$-electrons. 
We consider the nondegenerate case, i.e., one $d$- and one $f$-orbital per site 
is assumed. Therefore, owing to the two possible orientations of the spin, the 
average number of $d$- and $f$-electrons per site, $n_d$ and $n_f$, respectively, 
can vary between zero and two. The filling will refer to the ratio of the total 
electron density per site ($n_d + n_f$) and the maximally allowed electron density 
($n_{\rm max}=4$).  

Although it has been investigated for several decades,\cite{Review}  
this model and its extended versions are still in the forefront of 
condensed-matter physics. Since exact results are available only for certain 
special cases,\cite{Exact} besides the large number of perturbative studies 
nonperturbative techniques have also been developed to go beyond the 
weak-coupling limit. The Gutzwiller variational method\cite{Gutzwiller:original} 
has been applied by several authors.\cite{Rice&Ueda:variational,Brandow:1986,%
Varma:1986,Shiba:1986,Oguchi:1987,Fazekas:variational,Lamba:variational} 
In this method, an uncontrolled approximation (the so-called Gutzwiller 
approximation\cite{Gutzwiller:original}) is often used to calculate 
expectation values with the correlated wave function. Metzner and 
Vollhardt\cite{Vollhardt:cikk87} have shown that the expectation values
can be evaluated exactly in one dimension. Later they considered the 
limit of large dimensions,\cite{Vollhardt:cikk89} where 
analytic treatment is possible. Gebhard\cite{Gebhard:cikk} developed a 
technique to calculate expectation values in a controlled expansion in  
the inverse of the degeneracy of the $f$-level and in the inverse of the 
dimension of the lattice. He showed that the Gutzwiller approximation 
provides exact results in the limit of large 
dimensions. Moreover, in this limit this method is equivalent to the slave-boson 
mean-field theory of Kotliar and Ruckenstein.\cite{Kotliar:cikk,Moller:cikk} 
Later on, the dynamical mean-field theory,\cite{DMFT:review} which, too, is 
exact in the limit of infinite dimensions, has been applied to the periodic 
Anderson model by several authors\cite{Ohkawa,Jarrell1,Jarrell2,Saso} to better 
understand the main features of the model. To avoid the problem related to the 
Gutzwiller approximation, Shiba\cite{Shiba:1986} applied the variational 
Monte Carlo method. The PAM was investigated also by using the projector-based 
renormalization method\cite{Hubsch:cikk} for arbitrary degeneracy of the 
$f$-level. The ferromagnetic properties of the PAM have been studied with the 
density-matrix renormalization group.\cite{Guerrero:DMRG}

In view of the widespread application of the Gutzwiller approximation, it is 
important to know how reliable this method is. As will be demonstrated in 
this paper by comparing the results with those of exact 
diagonalization, the Gutzwiller method gives -- in spite of its limitations
-- reliable results for the number of electrons occupying the $f$-orbital.
The $f$-level occupancy is a significant quantity, for it has recently been
proposed\cite{Miyake_cikk} and experimentally verified\cite{RueffPRL_meres:cikk} 
that the pressure induced enhancement of the superconducting transition 
temperature of Ce based compounds, CeCu$_2$(Ge,Si)$_2$ is closely related to 
a sharp change of the valence of Ce.

Several extensions of the periodic Anderson model have been considered so 
far in order to make the model more realistic. It was found that 
nearest-neighbor interaction between $f$-electrons affects the stability 
of the magnetic ground state in the Kondo regime.\cite{Lamba:NN_int}
On the other hand, the on-site interaction between $d$- and $f$-electrons 
($U_{df}\sum_{\boldsymbol{j},\sigma,\sigma'}
\hat{n}^f_{j\sigma}\hat{n}^d_{j\sigma'}$)
influences drastically the occupation number of 
$f$-electrons.\cite{Hirashima:DMFT_Udf} It has been 
shown that a large $U_{df}$ destroys the Kondo state and narrows 
the intermediate valence regime.\cite{Miyake_cikk,Hirashima:DMFT_Udf} Its 
treatment in the framework of the Gutzwiller method is, however, quite 
cumbersome. In our previous work \cite{Hagymasi&Itai&Solyom:cikk} we 
assumed a special form for this interaction, $\widetilde{U}_{df}\sum_{\boldsymbol{j}}
\hat{n}^f_{j\uparrow}\hat{n}^f_{j\downarrow}
\hat{n}^d_{j\uparrow}\hat{n}^d_{j\downarrow}$, and 
pointed out that the intermediate-valence regime is narrowed in the presence 
of this interaction. 

The model we study in the second part of this paper includes the interaction 
between conduction electrons ($d$-electrons). Although the corresponding 
impurity problem has been examined thoroughly in several 
papers,\cite{Furusaki:cikk,Li:cikk,Frojd:cikk,%
Schork:cikk,Khaliullin:cikk,Igarashi:cikk,Igarashi:cikk1,Schork:cikk1}  
only a few results are available on the lattice problem.%
\cite{Itai:variational,Koga:DMFT_Ud,Schork:DMFT_Ud,Kawakami:Ud} 
Fulde and coworkers\cite{Fulde93} have pointed out that the heavy-fermion 
properties\cite{Brugger:exp} of Ce-doped Nd$_2$CuO$_4$ cannot be explained 
without taking correlations between conduction electrons into account. 
Although it has been shown\cite{Itai:variational} that correlations between 
conduction electrons may increase the effective mass, and the competition between 
Coulomb repulsion in the $d$- and $f$-electron subsystem may lead to a transition 
from Kondo to Mott insulator, the role of the electron-electron interaction 
in the conduction electron subsystem is not fully clarified. In this paper 
we calculate the number of $f$-electrons per site and the probability of 
double occupancy of $f$-orbitals as a function of the energy of the bare 
$f$-level, the hybridization, and the $f$-$f$ and $d$-$d$ Coulomb 
interactions. The calculations are carried out for a wide range of 
parameters of the model Hamiltonian, and the regions for Kondo-like 
behavior as well as for valence fluctuations are determined. 

The paper is divided into two main parts. Firstly, we investigate the 
reliability of the Gutzwiller method. We compare the variational results with 
those of exact diagonalization on finite chains. Secondly, we analyze 
what happens when the interaction between conduction electrons, $U_d$, is 
switched on.

\section{Variational calculation and exact diagonalization}

\subsection{Variational calculation}

First of all, following Ref.\ [\onlinecite{Itai:variational}] we summarize 
briefly the main steps of the variational calculation for the original
periodic Anderson model without interaction between conduction electrons, $U_d=0$.
In this paper we restrict ourselves to the paramagnetic case, i.e., the number of 
up-spin electrons, $N_{\uparrow}$, equals that of down-spin electrons, 
$N_{\downarrow}$.
Furthermore, we carry out the explicite calculation only for the system being half-filled or less than that, 
since the results for the system more than half-filled can be obtained straightforwardly owing to the electron-hole 
symmetry.

The trial wave function is chosen in the form
\begin{gather} 
\label{eq:variational_ansatz}
 |\Psi\rangle=\hat{P}_{\rm G}^f\prod_{\boldsymbol{k}}\prod_{\sigma}
  \left[u_{\boldsymbol{k}} \hat{f}_{\boldsymbol{k}\sigma}^{\dagger}
  +v_{\boldsymbol{k}} \hat{d}_{\boldsymbol{k}\sigma}^{\dagger}\right]|0\rangle,
\end{gather}
where the mixing amplitudes $u_{\boldsymbol{k}}$ and $v_{\boldsymbol{k}}$ 
are variational parameters. $\hat{P}_{\rm G}^f$ is the Gutzwiller projector for 
$f$-elect\-rons:
\begin{gather}
 \hat{P}_{\rm G}^f=\prod_{\boldsymbol{j}}\left[1-(1-\eta_f)
 \hat{n}_{j\uparrow}^f \hat{n}_{j\downarrow}^f \right],
\end{gather}
where the variational parameter $\eta_f$ is controlled by $U_f$. We use the 
Gutzwiller approximation to evaluate the expectation values. Optimizing with 
respect to the mixing amplitudes, we obtain 
\begin{gather}
 \mathcal{E}=\frac{1}{N}\sum_{\boldsymbol{k}\in 
   \mathrm{FS}}\left[\varepsilon_d(\boldsymbol{k})+\tilde{\varepsilon}_f    
   -\sqrt{\big[\varepsilon_d(\boldsymbol{k})-\tilde{\varepsilon}_f\big]^2+
    4\tilde{V}^2}\right]   \nonumber\\
      +(\varepsilon_f-\tilde{\varepsilon}_f)n_f+U_f\nu_f
\label{eq:energy1}
\end{gather}
for the ground-state energy per site, where $n_{f}$ and $\nu_f$ denote the 
number of $f$-electrons per site and the density of doubly occupied $f$-sites, 
respectively, $\tilde{V}=V\sqrt{q_f}$ is the renormalized hybridization 
amplitude with
\begin{eqnarray}
    q_f & = & \frac{1}{\left(1-\frac{n_f}{2}\right)\frac{n_f}{2}} 
    \Bigg[\sqrt{\left(\frac{n_f}{2} -\nu_f\right) \nu_f} \nonumber\\
     & & \phantom{+++} + \sqrt{\left(\frac{n_f}{2}-\nu_f\right) (1-n_f+\nu_f)}\Bigg]^2,
\label{q_f}
\end{eqnarray}
while the renormalized energy of the $f$-level, $\tilde{\varepsilon}_f$ has to be 
determined self-consistently from the condition 
\begin{gather}
   n_f = \frac{1}{N}\sum_{\boldsymbol{k}\in \mathrm{FS}} 
   \left[1 + \frac{\varepsilon_d(\boldsymbol{k})-\tilde{\varepsilon}_f}
   {\sqrt{\big[\varepsilon_d(\boldsymbol{k})-\tilde{\varepsilon}_f\big]^2 + 4
   {\tilde V}^2}}\right].
\label{eq:self-cons}
\end{gather}
The $\boldsymbol{k}$ sum in Eqs.\ (\ref{eq:energy1}) and 
(\ref{eq:self-cons}) [and later on, Eqs.\ \eqref{eq:energy} and \eqref{eq:self-cons2} in the next section] extends over the $U_f = U_d = 0$ Fermi 
sea in a manner familiar from the periodic Anderson 
model,\cite{Fazekas:variational} since the Gutzwiller method 
respects Luttinger's theorem and leaves the Fermi volume unchanged.

The quantities $n_f$ and $\nu_f$, and thereby $\tilde{\varepsilon}_f$ and $q_f$
depend on the as yet undetermined variational parameter $\eta_f$. Optimizing
with respect to this parameter is equivalent to minimizing the energy
with respect to $n_f$ and $\nu_f$, which leads to 
\begin{eqnarray}
{\tilde\varepsilon}_f &=& 
\frac{\partial{\cal E}}{\partial q_f}\cdot\frac{\partial q_f}{\partial n_f} + 
\frac{\partial{\cal E}}{\partial {\tilde\varepsilon}_f} \cdot\frac{\partial 
{\tilde\varepsilon}_f}{\partial n_f},
\label{eq:shift}\\
- U_f &=&  \frac{\partial{\cal E}}{\partial q_f}\cdot\frac{\partial q_f}
{\partial {\nu_f}} + \frac{\partial{\cal E}}{\partial {\tilde\varepsilon}_f} 
\cdot\frac{\partial {\tilde\varepsilon}_f}{\partial \nu_f}.
\label{eq:nuf}
\end{eqnarray}
These equations have to be solved together with the self-consistency condition 
(\ref{eq:self-cons}).
 
The summation over $\boldsymbol{k}$ in Eqs.\ (\ref{eq:energy1}) and 
(\ref{eq:self-cons}) could be carried out numerically 
for a realistic dispersion curve $\varepsilon_d(\boldsymbol{k})$, but 
the variational procedure, i.e., the numerical optimization of the 
ground-state energy with the self-consistency condition (\ref{eq:self-cons}), 
would be very cumbersome. Instead of that, we assume a constant density of 
states, $\rho(\varepsilon)=1/W$, in the interval $\varepsilon\in[-W/2,W/2]$, 
since then the energy density and the self-consistent value of 
$\tilde\varepsilon_f$ can be expressed analytically from Eqs.\ \eqref{eq:energy1} 
and \eqref{eq:self-cons} as a function of $n_f$ and $\nu_f$. 

However, the self-consistent solution of the minimum conditions for $n_f$ 
and $\nu_f$ can be found analytically only in special cases, e.g., 
for $U_f \rightarrow \infty$, when $V \ll W$.\cite{Itai:variational}  
In this paper we will solve Eqs.\ (\ref{eq:shift}) and 
(\ref{eq:nuf}) numerically for various values of $V$, $U_f$, and $\varepsilon_f$ 
in order to determine the range of parameters for the Kondo or 
intermediate-valence behavior, and for the crossover regime between them.

Firstly, we calculate the $U_f$- and $\varepsilon_f$-dependence of the 
$f$-level occupancy, $n_f$, and of the renormalization factor $q_f$ in the 
half-filled case, where the total number of electrons equals the sum of 
the number of $d$- and $f$-orbitals (the electron density per site 
$n = n_d + n_f = 2$), and in the 1/3-filled case ($n=4/3$). Other fillings 
will be discussed later in the next subsection, where 
we compare the results with those obtained by exact diagonalization.

We note  that our model with $n$ electrons can be mapped onto a model with 
$n$ holes ($4-n$ electrons), provided that the energy level of the $f$-hole 
is chosen as $-(\varepsilon_f + U_f)$. Therefore, the results for $n > 2$ can 
be obtained straightforwardly from those for $n < 2$. Owing to this symmetry 
in the special, symmetric half-filled case, when $n\!=\!2$ and the bare $f$-level
is located at $\varepsilon_f=-U_f/2$, both $n_f$ and $n_d$ are exactly equal to 1. 

\begin{figure}[!htb]
\includegraphics[scale=0.5]{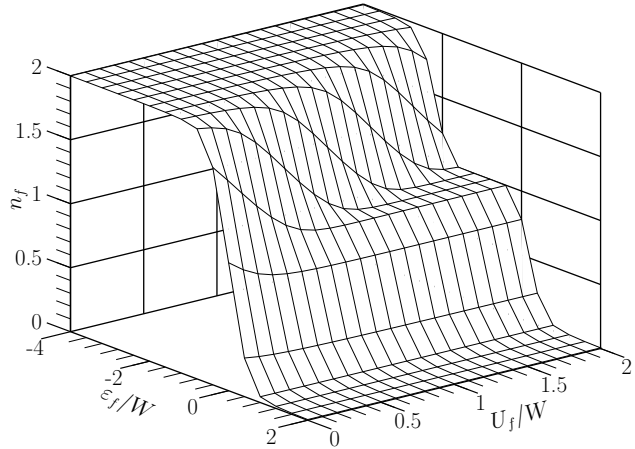}
\caption{\label{nf_Uf_plato:fig}
The $f$-level occupancy as a function of $\varepsilon_f$ 
and $U_f$ at half filling ($n=2$) for $V/W=0.1$.}
\end{figure}

\begin{figure}[!ht]
\includegraphics[scale=0.5]{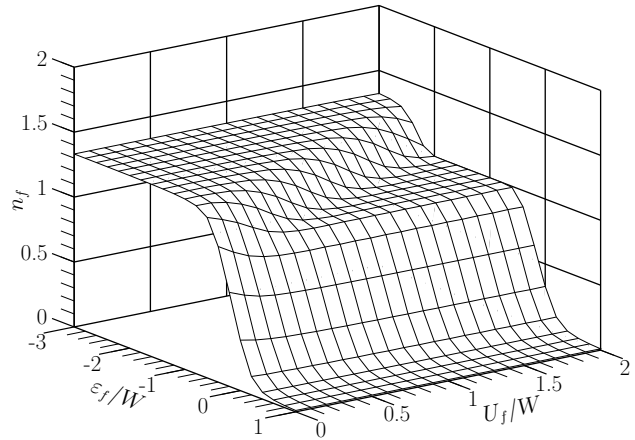}
\caption{\label{nf_Uf_plato-8hatod:fig}
The $f$-level occupancy as a function of $\varepsilon_f$ 
and $U_f$ at 1/3 filling ($n=4/3$) for $V/W=0.1$.}
\end{figure}

The $f$-level occupancy is displayed as a function 
of the bare $f$-level energy and of $U_f$ for $V/W = 0.1$ in the half-filled 
and 1/3-filled cases, respectively, in Figs.\ \ref{nf_Uf_plato:fig} and 
\ref{nf_Uf_plato-8hatod:fig}. Five different regimes can be distinguished. When 
$\varepsilon_f+U_f$ lies below the conduction band, all electrons occupy 
$f$-orbitals, $n_f \approx 2$ and $4/3$, respectively. The regions, where 
$n_f$ varies smoothly, almost linearly from 2 (or 4/3) to 1 and later from 1 
to 0, are the intermediate-valence regimes. On the plateau between them, 
$n_f$ deviates from unity by an exponentially small amount. This is, 
as we will see, the Kondo regime, since the double-occupancy rate is 
exponentially small here. Finally, when $\varepsilon_f$ lies well above 
the conduction band, all electrons occupy states in the conduction band, 
and $n_f\approx 0$. There are no sharp boundaries between these regimes; 
narrow crossover regions separate them. 

\begin{figure}[!htb]
\includegraphics[scale=0.26]{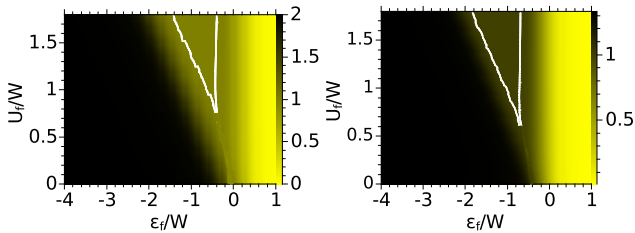}
\caption{\label{phase-diag-n_f:fig} (Color online)
The $f$-level occupancy as a function of $\varepsilon_f$ 
and $U_f$ at half filling (left) and 1/3 filling (right) for $V/W=0.1$. 
The color code is shown at the right edge of the panels. The boundary of 
the $n_f\approx1$ plateau is drawn with a white line.}
\end{figure}

The boundary of the $n_f \approx  1$ plateau could be defined by 
setting a somewhat arbitrary criterion for the deviation of $n_f$ 
from unity. Figure \ref{phase-diag-n_f:fig} shows 
$n_f$ in the $U_f$--$\varepsilon_f$ plane for a particular value of $V/W$ 
using a color code. The ``boundary'' of the plateau defined by 
$|1-n_f|=0.005$ is drawn with a white line. As can be seen in the figures, 
a plateau develops only when $U_f$ exceeds a not sharply 
defined threshold value, $U^{\text{c}}_{f}$, which itself depends on $V$ 
and on the total electron density. Besides $V/W = 0.1$, we have done 
calculations for $V/W = 0.05$ and $0.2$, and obtained similar results. The 
upper and lower limits of $\varepsilon_f$ between which the plateau forms 
can be estimated from the numerical data to be roughly
\begin{gather}
 -U_f + E_{\rm F}(n_d) + a\Delta_f \lesssim\varepsilon_f \lesssim E_{\rm F}(n_d) - a\Delta_f, 
\label{eq:plateau}
\end{gather}
where $E_{\rm F}(n_d)$ is the Fermi level of the conduction band with 
$n_d\!=\!n\!-\!1$ electrons, $\Delta_f = \pi\rho V^2$ is the width of the $f$-level 
in the impurity problem, and $a$ is a numerical factor of order 10, which 
depends weakly on $V$, $U_f$, and $n$. The factor $a$ is smaller by about 
$10\%$ for $n=4/3$ than for $n=2$, which shows that the plateau slightly expands as the 
filling of the conduction band decreases from half filling. 

These results are somewhat surprising. One could argue, based on the results 
for the impurity Anderson model that a Kondo-like behavior (i.e., $n_f \approx 
1$ with very small valence fluctuations) is realized when the Fermi level is 
located between the bare $f$-level ($\varepsilon_f$) and the energy 
$\varepsilon_f + U_f$ of a second $f$-electron occupying the same site. 
That is, we could expect the condition $-U_f+E_{\rm F} \lesssim 
\varepsilon_f \lesssim E_{\rm F}$, when $\Delta_f \ll W$. Condition 
(\ref{eq:plateau}) obtained by the Gutzwiller method indicates that 
the Kondo-like behavior is realized in the periodic Anderson model in a
much narrower interval for $\varepsilon_f$. This will be confirmed later 
by exact diagonalization.

\begin{figure}[!htb]
\includegraphics[scale=0.5]{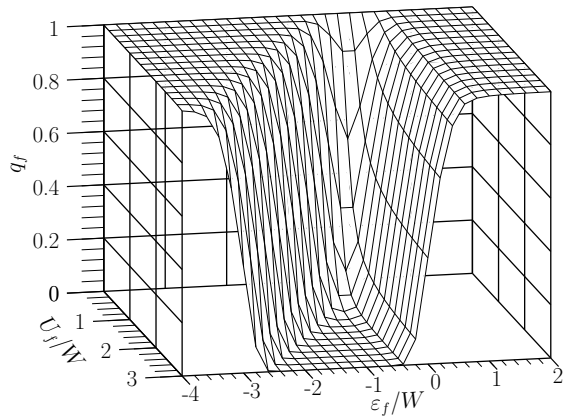}
\caption{\label{qf_Uf:fig}
The kinetic energy renormalization factor of $f$-elec\-trons as a function 
of $\varepsilon_f$ and $U_f$ at half filling for $V/W=0.1$.}
\end{figure}

The $f$-electrons are strongly correlated on this plateau, since not only 
the average occupancy of the $f$-orbital is close to unity there, but the 
number of empty or doubly occupied $f$-orbitals is almost negligible. 
Correlations between $f$-electrons can conveniently be characterized 
by the renormalization factor $q_f$, which is simply related to the 
double-occupancy rate $\nu_f$ as
\begin{equation} \label{eq:qf_szimm_pont}
     q_f = 8\nu_f(1-2\nu_f),
\end{equation}
when $n_f$ is exactly one. This quantity is plotted versus $\varepsilon_f$ 
and $U_f$ in Fig.\ \ref{qf_Uf:fig} for $V/W=0.1$ at half filling. 

\begin{figure}[!htb]
\includegraphics[scale=0.26]{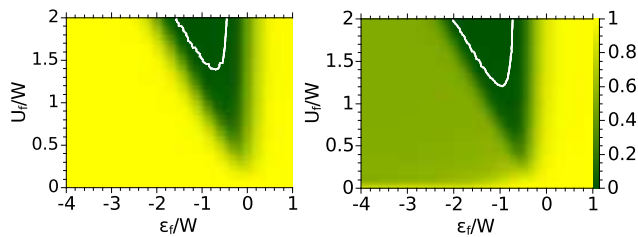}
\caption{\label{phase-diag} (Color online)
The parameter $q_f$ is displayed for $V/W = 0.1$ using a color code shown 
at the right edge. The boundary of the Kondo regime defined by $q_f = 0.005$ 
is drawn with a white line. Left: half-filled case, right: 1/3-filled case.}
\end{figure}

It is clearly seen that $q_f$ decreases rapidly from about $1$, when
the $f$-level is doubly occupied or empty, to about $0$ as $n_f$ approaches 
one from either side. When $q_f \approx 0$, the double-occupancy rate 
is also close to zero, and the $f$-electrons show heavy-fermion behavior; 
the effective mass becomes large as $m^*\propto q_f^{-1}$. We can, therefore, 
define the Kondo regime by setting a limit on $q_f$, by requiring, e.g., 
$q_f < 0.005$. This boundary is marked by a white line in Fig.\ \ref{phase-diag}, 
where $q_f$ is shown for $n=2$ and $n=4/3$ using a color code. 

The Kondo regime thus defined appears again above a critical $U_f^{\text{c}}$, 
which is, however, somewhat larger than the one found earlier, since
the criterion $|1- n_f| \leq 0.005$ is less strict than
the condition $q_f < 0.005$. In this latter case the probability of double 
occupancy has to be less than $0.0006$. Nevertheless, comparison with 
Fig.\ \ref{phase-diag-n_f:fig} shows that apart from
a rounding around the critical $U_f^{\text{c}}$, the two criteria define
the same regime. The plateau slightly expands when the filling of the 
conduction band  decreases.

When $n_f$ is exactly one, and Eq.\ (\ref{eq:qf_szimm_pont}) holds,
Eq.\ (\ref{eq:nuf}) can be easily solved in the limit $V \ll W$. We get 
\begin{gather}
q_f = \frac{n_d}{4(V/W)^2}\; \text{exp} \left( -\;\frac{U_f}{16V^2/W}\right),
\label{q_f-symm}
\end{gather}
where $n_d$ is the number of the conduction electrons per site. The factor 
$n_d$ in the prefactor explains why the critical $U^{\text{c}}_f$ gets 
smaller as the filling decreases. 

The total energy density takes a simple form in this limit, 
\begin{gather}
{\mathcal E} = \varepsilon_f + E_d(n_d) - n_d \frac{W}{2} \exp 
{\left\{-\frac{U_f}{16V^2/W}\right\}},
\label{eq:Esym}
\end{gather}
where the first term is the energy of the half-filled $f$-orbital without 
polarity fluctuations, $E_d(n_d)$ is the energy of the decoupled 
conduction band for filling $n_d$, and the last term describes the coupling 
between $f$-electrons and conduction electrons, in other words, the energy 
decrease owing to the polarity fluctuations caused by $d$-$f$ hybridization. 
This term arising from the Kondo effect gives the characteristic energy 
scale in the Kondo regime. The Kondo energy, $E_{\text K}$, is defined by the 
energy decrease per conduction electron, i.e.,
\begin{gather}
E_{\text K} = \frac{W}{2} \exp {\left\{-\frac{U_f}{16V^2/W}\right\}}.
\label{eq:kondo-scale}
\end{gather}

\begin{figure}[!ht]
\includegraphics[scale=0.5]{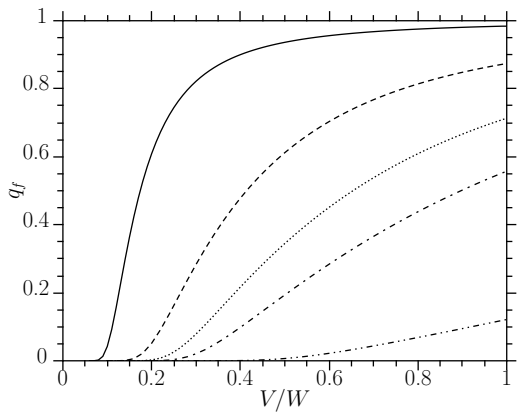}
\caption{\label{qf_V:fig}
$q_f$ vs.\ $V$ in the symmetric half-filled case, $\varepsilon_f = -U_f/2$, for
$U_f/W =1$ (continuous line), 3 (dashed), 
5 (dotted), 7 (dashed with one dot), and 20 (dashed with two dots), respectively.}
\end{figure}

In the remaining part of this subsection, we study more quantitatively the 
dependence of the threshold value of $U_f^{\text{c}}$ on $V$ in the 
half-filled case. Figure \ref{qf_V:fig} shows $q_f$ as a function of $V$ 
for several values of $U_f$ at $\varepsilon_f = -U_f/2$, where $n_f$ is 
exactly one.
The threshold values determined from $q_f(U^{\text{c}}_{f},V) = 0.005$ 
are given in Table \ref{tabla:J} together the corresponding Kondo coupling 
$J = 8V^2/U_f^{\text{c}}$, since, in the Kondo regime, the periodic Anderson 
model can be mapped onto a Kondo lattice model (KLM). 

\begin{table}[h]
\begin{tabular}{@{}c@{\hspace{4mm}}c@{\hspace{4mm}}c@{\hspace{4mm}}}
\toprule
$V/W$ &  $U^{\text{c}}_f/W$ &  $ J/W$\\ \hline
$0.16$ &  3 &  $0.066$\\
$0.21$ &  5 &  $0.071$\\
$0.26$ &  7 &  $0.076$\\
$0.48$ &  20 &  $0.093$\\
$0.74$ &  40 & $0.110$\\
$0.97$ &  60 &  $0.125$\\
$1.17$ &  80 &  $0.136$\\
$1.36$ &  100 &  $0.148$\\
\toprule
\end{tabular}
\caption{The critical $U^{\text{c}}_{f}$ of the Kondo plateau for
several values of $V$ and the corresponding Kondo coupling.}
\label{tabla:J}
\end{table}
The dependence of $U^{\text{c}}_f$ on $V$ can 
be fitted by the analytic functional form
\begin{gather}
    U^{\text{c}}_f/W =62.56(V/W)^{\alpha} .
\label{eq:fit}
\end{gather}
with $\alpha = 1.54$

Since by definition there are no doubly occupied or vacant $f$-orbitals in 
a KLM, a rigorous mapping from PAM to KLM should be possible in the limit 
$\nu_f \rightarrow 0$. Setting a smaller limit for $q_f$ in the criterion 
for the Kondo regime, larger exponents, given in Table \ref{tabla:exponent}, 
and larger numerical prefactors are obtained in Eq.\ (\ref{eq:fit}). The 
exponent $\alpha$ seems to converge to 2 in 
the limit $q_f \rightarrow 0$, which means that $U_f^{\text{c}}$ is 
proportional to $\Delta_f$, and the proportionality factor is of order hundred
instead of the factor $a\approx 10$ in Eq.\ \eqref{eq:plateau}. This
difference is due to the stricter condition on $q$ and to the rounding of
the boundary at the critical $U_f^{\text{c}}$.

\begin{table}[h]
\begin{tabular}{@{}c@{\hspace{3mm}}c@{\hspace{3mm}}c@{\hspace{3mm}}c
@{\hspace{3mm}}c@{\hspace{3mm}}c@{\hspace{3mm}}c@{}}
\toprule
$q_f^{\text{threshold}}$ & $10^{-3}$ & $10^{-4}$ & $10^{-5}$ & $10^{-6}$ 
  & $10^{-7}$ & $10^{-8}$\\
\hline
$\alpha$ & 1.70 & 1.80 & 1.83 & 1.86 & 1.89 & 1.91\\
\toprule
\end{tabular}
\caption{The exponent $\alpha$ in Eq.\ \eqref{eq:fit} calculated for several 
threshold value of $q_f$.}
\label{tabla:exponent}
\end{table}

Sinjukow and Nolting\cite{Nolting:2002} have shown that in the extended 
Kondo limit, when $U_f \rightarrow \infty$ and $V \rightarrow \infty$ with 
$V^2/U_f$ remaining finite, the symmetric periodic Anderson model 
can be mapped exactly to the Kondo lattice model with finite Kondo coupling. 
The results obtained by the Gutzwiller method are in agreement with this.

\subsection{Comparison with  exact diagonalization}

With the aim to compare the variational results with those of a completely 
different method, we also performed exact diagonalization 
on relatively short chains. In order to check whether the results 
obtained for these chains are representative for bulk materials, we 
calculated the $f$-level occupancy, $n_f$, and the density of doubly occupied 
$f$-sites, $\nu_f$, in the nonmagnetic ($S_z^{\rm tot}=0$) ground state 
for chains of four, five, and six sites. It turned out that the results were 
in excellent agreement with each other, which suggests that the six-site chain 
behaves almost like the bulk in this respect. This is in agreement with the
finding of Chen and Callaway,\cite{Callaway:diag&MC} who compared the 
ground-state energy obtained from exact diagonalization of a four-site 
chain with Monte Carlo result on a sixteen-site chain. In what follows we 
present the results obtained for a six-site chain with 12, 10, 8, and 6 electrons.
The case with 6 electrons is not interesting  from the point of view of Kondo 
physics, because the conduction band is exhausted when $n_f=1$. Nevertheless,
it is used in the comparison of the two methods. 

The kinetic energy of conduction electrons moving along the chain is described 
by hopping between nearest-neighbor $d$-orbitals with hopping rate $t$, thus 
the band width is now $4t$. Therefore, we identify $W$ with $4t$, when comparison
with the results of the variational calculation is made. 

\begin{figure}[!ht]
\includegraphics[scale=0.5]{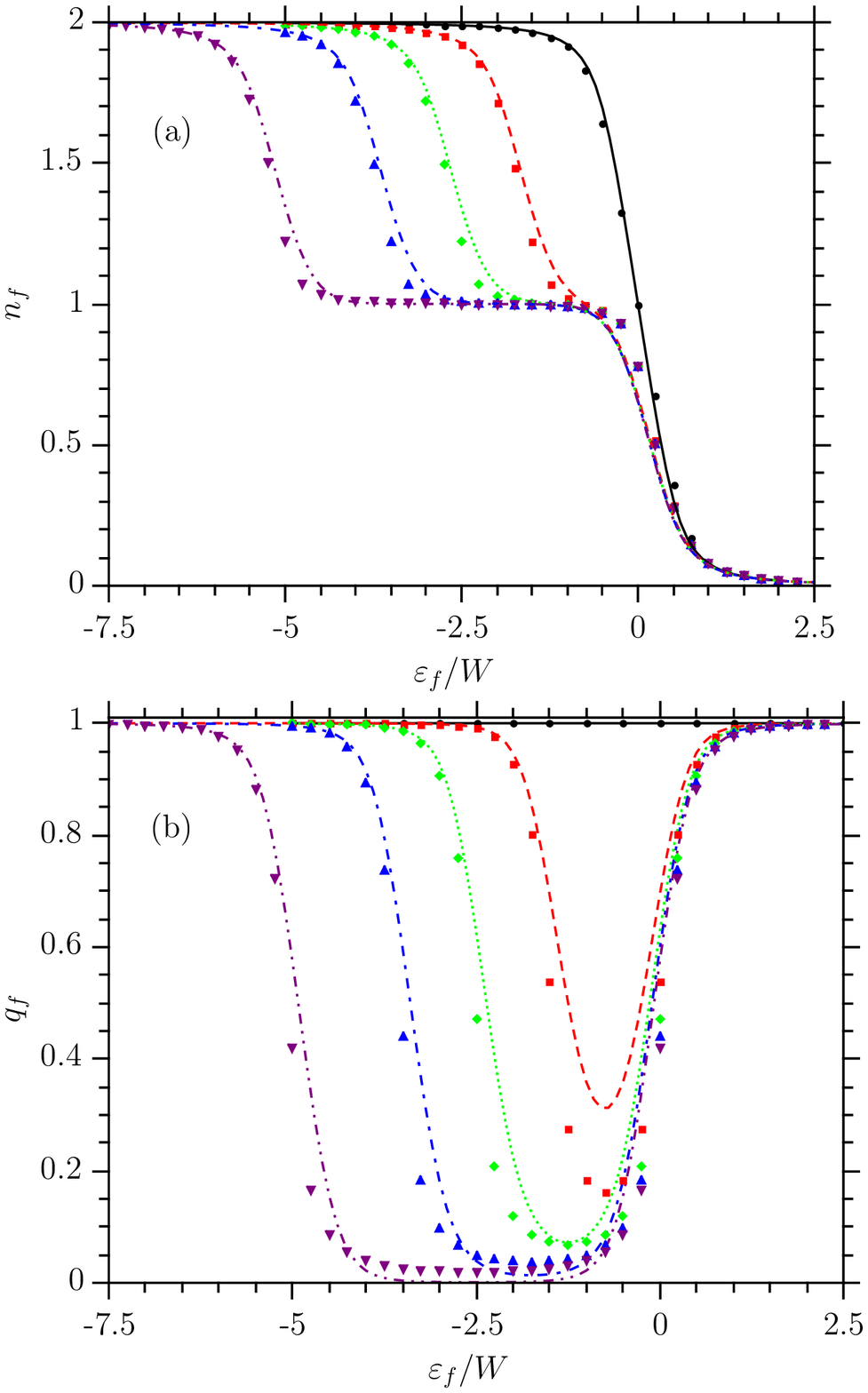}
\caption{\label{nf_ex&GW:fig}
(Color online) (a) The $f$-level occupancy vs.\ $\varepsilon_f$
at $2V/W = 0.375$. The curves are obtained by the Gutzwiller method, while 
the symbols indicate the results of exact diagonalization for $2U_f/W =0$ 
(black continuous line, {\tiny\textbullet}), 3 (red dashed line, 
{\tiny{$\blacksquare$}}), 5 (green dotted line, {\tiny{$\blacklozenge$}}), 
7 (blue dashed line with one dot, {\tiny{$\blacktriangle$}}) and 10 
(purple dashed line with two dots, {\tiny{$\blacktriangledown$}}). 
(b) The renormalization factor $q_f$. The notation 
is the same as in panel (a).}
\end{figure}

The $f$-level occupancy obtained by the two methods are directly compared 
in  Fig. \ref{nf_ex&GW:fig}(a). As for $\nu_f$, we compare the results indirectly, 
through $q_f$. Although this quantity is specific to the Gutzwiller method, it 
shows the strength of correlations more visibly than $\nu_f$ itself, therefore, 
we define $q_f$ with the help of Eq.\ (\ref{q_f}) from $n_f$ and $\nu_f$ 
obtained from the exact ground-state wave function. Comparison with the result 
of the variational calculation is shown in Fig.\ \ref{nf_ex&GW:fig}(b).

As is seen in Fig.\ \ref{nf_ex&GW:fig}(a), the two methods give very similar 
results as far as the ``global behavior'' of the $f$-level occupancy and the 
extent of the $n_f \approx 1$ plateau is concerned, even though the density 
of states is not identical in the two calculations. This indicates that 
Eq.\ (\ref{eq:plateau}) found in the Gutzwiller method for the boundary of 
the Kondo regime is not due to the Gutzwiller approximation, but is a 
consequence of strong correlations in the lattice model.

We find a subtle difference, however, in Fig.\ \ref{nf_ex&GW:fig}(b), where 
$q_f$ is plotted as a function of $\varepsilon_f$. One sees that $q_f$ 
calculated in the Gutzwiller method approaches zero faster in the Kondo 
regime than that provided by exact diagonalization. The former exhibits 
the exponential behavior given in Eqs.\ (\ref{q_f-symm}) and 
(\ref{eq:kondo-scale}) typical for Kondo physics, while the latter cannot be 
fitted to such a curve. We will discuss this quantitatively later on. 

\begin{figure}[!b]
\includegraphics[scale=0.5]{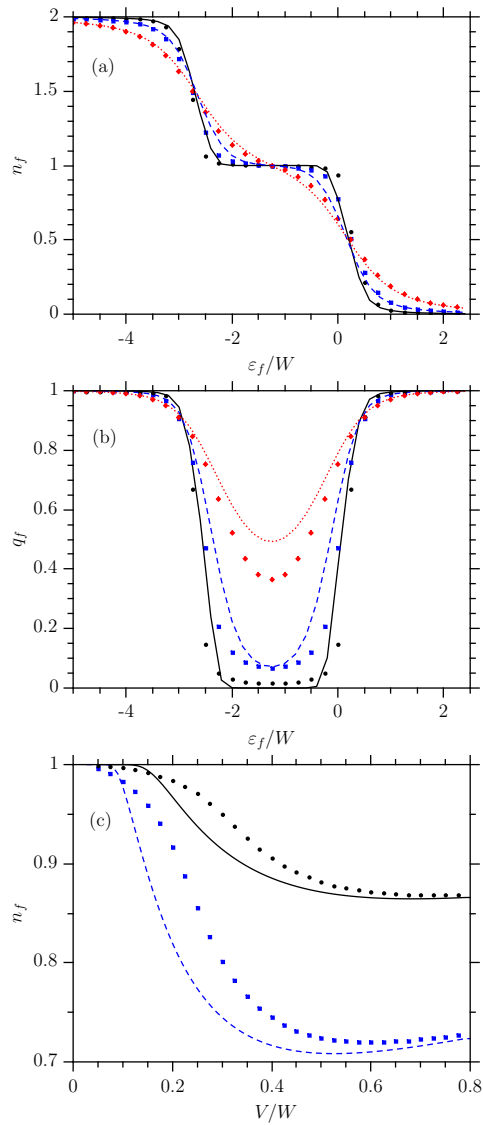}
\caption{\label{nf_qf_V:fig}
(Color online) 
(a) The $f$-level occupancy vs.\ $\varepsilon_f$
at $U_f/W=2.5$. The curves are obtained by the Gutzwiller method, while
symbols denote the results of exact diagonalization for $2V/W=0.2$ 
(black continuous, {\tiny\textbullet}), $0.375$ (blue dashed, 
{\tiny{$\blacklozenge$}}), and $0.7$ (red dotted, {\tiny{$\blacksquare$}}), 
respectively. (b) The renormalization factor $q_f$. 
The notation is the same as in panel (a). (c) $n_f$ vs.\ $V$ at $U_f/W = 2.5$, 
for $\varepsilon_f /W =-0.75$ (black continuous line, {\tiny\textbullet}) 
and $\varepsilon_f /W =-0.25$ (blue dashed line, {\tiny{$\blacksquare$}}).}
\end{figure}

Next we check the dependence of the Kondo plateau on the strength of the
hybridization. In Fig.\ \ref{nf_qf_V:fig}(a) we plot $n_f$ as a function of
$\varepsilon_f$ for three values of $V/W$ in the half-filled case.
It is clearly seen that the plateau (i.e., the Kondo regime) 
rapidly shrinks as $V$ increases, and disappears, in agreement with the
results presented in the previous subsection. Figure \ref{nf_qf_V:fig}(b), 
where $q_f$ is plotted, shows directly the disappearance of heavy-fermion 
behavior. Finally $n_f$ is plotted as a function of $V$ in Fig.\ 
\ref{nf_qf_V:fig}(c) for two values of $\varepsilon_f/W$. We find again 
that the two methods yield similar results for $n_f$, but the 
$V$-dependence is different near the boundary of the Kondo regime.

\begin{figure}[!t]
\includegraphics[scale=0.33]{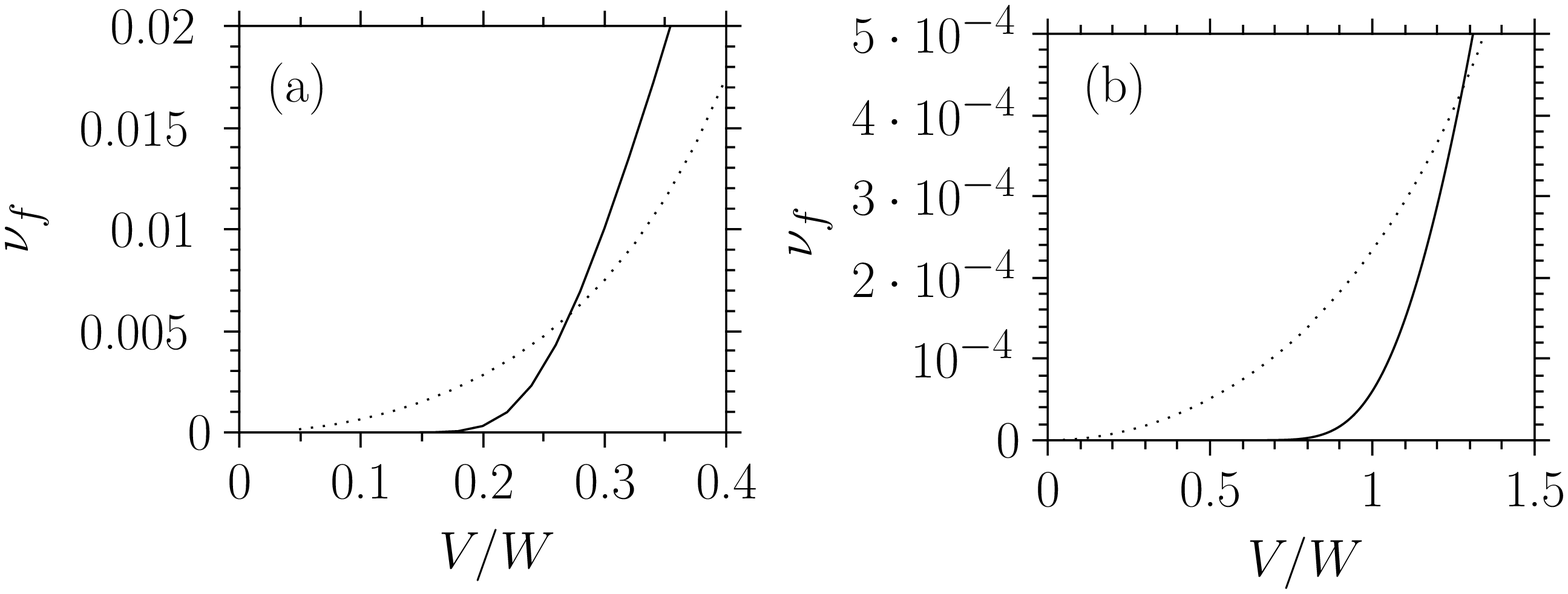}
\caption{\label{nuf_V:fig}
The double-occupancy rate of $f$-electrons vs.\ $V/W$. The dotted 
curves indicate the results of exact diagonalization, while the continuous 
curves are calculated by the Gutzwiller method. $U_f/W$ is 5 and 100 
in panel (a) and (b), respectively. }
\end{figure}

In order to better see this difference, we calculate the double-occupancy rate of 
$f$-electrons in the symmetric ($\varepsilon_f=-U_f/2$) half-filled case as a 
function of $V$ near the boundary of the Kondo regime, i.e., where $\nu_f \ll 1$.
We find, as seen in Fig. \ref{nuf_V:fig}, that in contrast to the results of the 
Gutzwiller method, the dependence of $\nu_f$ on $V^2/U_f$ is not
exponential; $\nu_f$ varies as a power of $V^2/U_f$: 
\begin{eqnarray}
\nu_f = A\; \frac{W}{U_f}\left(\frac{V^2}{WU_f}\right) + B\; 
\frac{W}{U_f}\left(\frac{V^2}{WU_f}\right)^2,
\label{qf_V:ED}
\end{eqnarray}
where $A$ is close to unity and $B \approx 50$.
This power-law-like dependence may be due to the small system size in the 
exact diagonalization. 

\begin{figure}[!htb]
\includegraphics[scale=0.5]{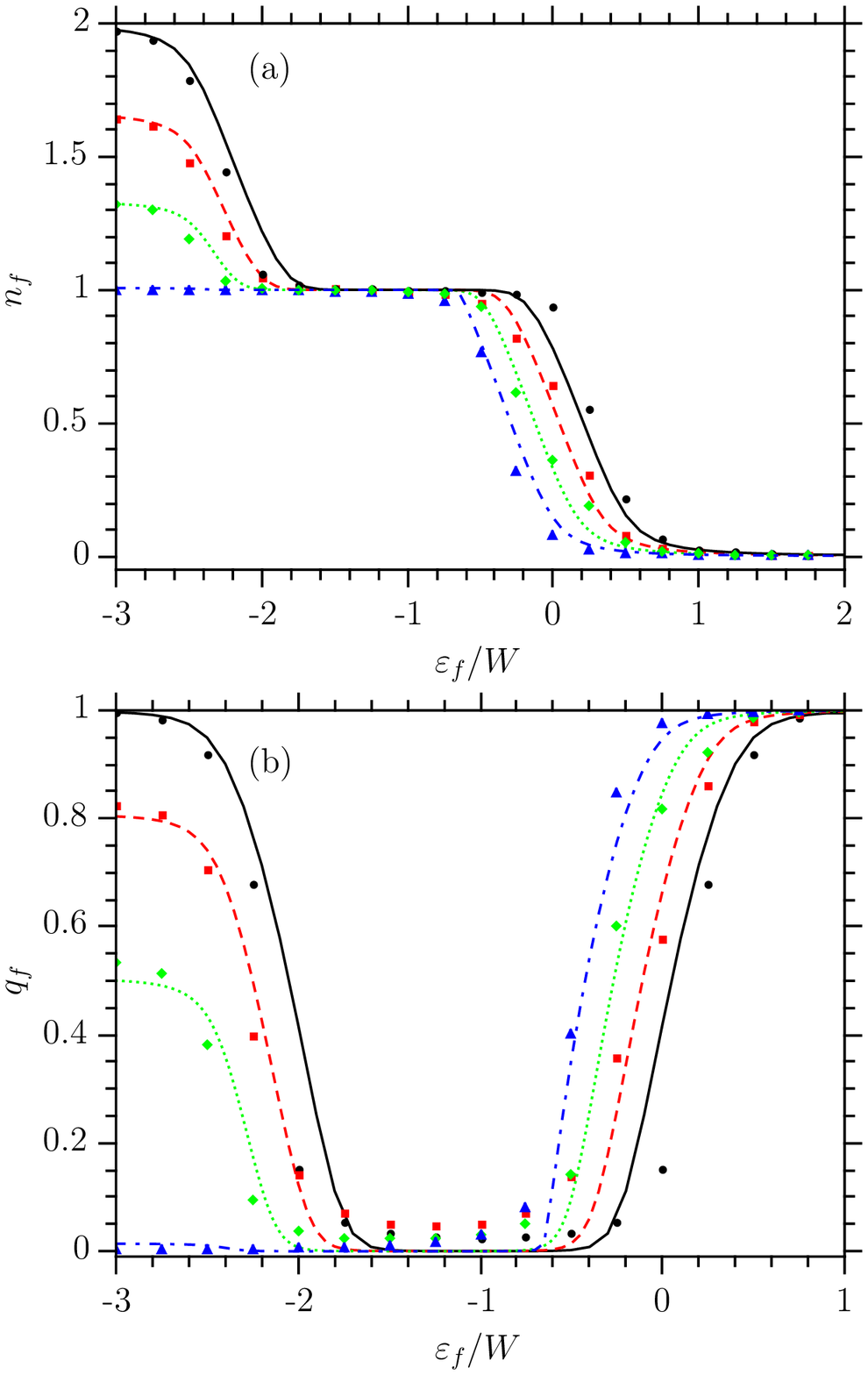}
\caption{\label{filling1:fig} (Color online) (a) The $f$-level occupancy vs.\
$\varepsilon_f$ at $U_f/W=2$ for different fillings. The 
hybridization is $V/W = 0.1$ in all cases. The curves are obtained by the 
Gutzwiller method, while the symbols are the results of exact diagonalization. 
The number of electrons per site is $n=2$ (black continuous line, 
{\tiny\textbullet}), 5/3 (red dashed line, {\tiny{$\blacksquare$}}), 4/3 
(green dotted line, {\tiny{$\blacklozenge$}}), 1 (blue dashed line with one dot,
{\tiny{$\blacktriangle$}}), respectively. (b) The $f$-level kinetic energy 
renormalization factor. The notation is the same as in panel (a).}
\end{figure}

Finally we study the filling dependence of the Kondo regime. The $f$-level 
occupancy is shown for several fillings in Fig.\ \ref{filling1:fig}(a). The 
overall agreement between the two methods persists as we move away from 
half filling, though its degree varies somewhat, e.g., the agreement 
in the $n=2$ or 5/3 case is noticeably less good than for $n=4/3$. This 
indicates that the Gutzwiller type paramagnetic wave function is more 
appropriate for metallic systems with few conduction electrons than for 
insulators. The Kondo plateau shifts towards lower $f$-level energies 
as the filling decreases owing to the decrease of the Fermi level. We find 
a similar slight difference between the results of the two methods 
displayed in Fig.\ \ref{filling1:fig}(b), 
where $q_f$ is plotted as a function of $\varepsilon_f$.

\section{The role of interaction between conduction electrons}

\subsection{Variational calculation}

As a next step, we consider what happens when the interaction between conduction 
electrons is switched on. For the sake of simplicity a local, on-site interaction is 
assumed and the Hamiltonian takes the form
\begin{equation}  
\mathcal{H} = \mathcal{H}_{\textrm{PAM}} + 
     U_d\sum_{j}\hat{n}^d_{j\uparrow}
   \hat{n}^d_{j\downarrow}\;,
\end{equation}
where $\mathcal{H}_{\textrm{PAM}}$ is the PAM Hamiltonian defined in Eq.\ 
(\ref{PAM:Hamiltonian}) and $U_d$ is the strength of the Coulomb interaction 
between conduction electrons. This model is also known as the periodic 
Anderson-Hubbard model. At half filling the symmetric model corresponds to 
$\varepsilon_f = -U_f/2 + U_d/2$, where $n_f = n_d = 1$.

The variational calculation can be performed by a simple generalization of 
the procedure used for $U_f \rightarrow \infty$.\cite{Itai:variational}
The trial wave function is chosen in the form
\begin{gather} \label{eq:variational_ansatz_Ud}
 |\Psi\rangle=\hat{P}_{\text{G}}^f\hat{P}_{\text{G}}^d\prod_{\boldsymbol{k}}
    \prod_{\sigma}\left[u_{\boldsymbol{k}}
 \hat{f}_{\boldsymbol{k}\sigma}^{\dagger}+v_{\boldsymbol{k}}
 \hat{d}_{\boldsymbol{k}\sigma}^{\dagger}\right]|0\rangle,
\end{gather}
where $\hat{P}^f_{\text{G}}$ contains the variational parameter $\eta_f$, and an 
extra Gutzwiller projector has been introduced for $d$-electrons, 
which is written as 
\begin{gather}
 \hat{P}_{\text{G}}^d=\prod_{\boldsymbol{j}}\left[1-(1-\eta_d)
    \hat{n}_{j\uparrow}^d\hat{n}_{j\downarrow}^d\right].
\end{gather}
The variational parameter $\eta_d$ depends on $U_d$. Performing 
the optimization with respect to the mixing amplitudes we get 
\begin{gather}
 \mathcal{E}=\frac{1}{N}\sum_{\boldsymbol{k}\in    
     \mathrm{FS}}\left[q_d\varepsilon_d(\boldsymbol{k})+
    \tilde{\varepsilon}_f-\sqrt{\big[q_d\varepsilon_d(\boldsymbol{k})
     -\tilde{\varepsilon}_f\big]^2+4\tilde{V}^2}\right]   \nonumber\\
      +(\varepsilon_f-\tilde{\varepsilon}_f)n_f+U_d\nu_d+U_f\nu_f
\label{eq:energy}
\end{gather}
for the ground-state energy density, where $\nu_d$ is the density of 
doubly occupied $d$-sites, and $q_d$ denotes the kinetic energy 
renormalization factor of $d$-electrons given by 
\begin{eqnarray}
    q_d  & = &  \frac{1}{\left(1-\frac{n_d}{2}\right)\frac{n_d}{2}} 
    \Bigg[\sqrt{\left(\frac{n_d}{2} -\nu_d\right) \nu_d} \nonumber\\
        & & + \sqrt{\left(\frac{n_d}{2}-\nu_d\right) (1-n_d +\nu_d)}\;\Bigg]^2,
\label{q_d}
\end{eqnarray}
which is formally identical to that found in the Hubbard model.\cite{Gutzwiller:original} 
The renormalized hybridization amplitude is now 
$\tilde{V}=V\sqrt{q_dq_f}$; the other notations are the same as 
in the previous section, and the self-consistency condition [see Eq.\ 
(\ref{eq:self-cons})] is now given by
\begin{gather}
   n_f = \frac{1}{N}\sum_{\boldsymbol{k}\in \mathrm{FS}} 
   \left[1 + \frac{q_d\varepsilon_d(\boldsymbol{k})-\tilde{\varepsilon}_f}
   {\sqrt{\big[q_d\varepsilon_d(\boldsymbol{k})-\tilde{\varepsilon}_f\big]^2 + 4
   {\tilde V}^2}}\right].
\label{eq:self-cons2}
\end{gather}
The summation over $\boldsymbol{k}$ and the numerical 
optimization of the energy density with respect to $n_f$, $\nu_f$, and $\nu_d$ 
are carried out in the same way as in the previous section. 
The equations determining $n_f$, $\nu_f$ and $\nu_d$ are now 
\begin{eqnarray}
{\tilde\varepsilon}_f & = & \frac{\partial{\cal E}}{\partial q_d}\cdot
   \frac{\partial q_d}{\partial n_f} + 
\frac{\partial{\cal E}}{\partial q_f}\cdot\frac{\partial q_f}{\partial n_f} +
\frac{\partial{\cal E}}{\partial {\tilde\varepsilon}_f}\cdot
   \frac{\partial {\tilde\varepsilon}_f}{\partial n_f},
\label{eq:shift2}\\
  - U_f & = & \frac{\partial{\cal E}}{\partial q_f}\cdot\frac{\partial q_f}
{\partial {\nu_f}} + \frac{\partial{\cal E}}{\partial {\tilde\varepsilon}_f} 
   \cdot\frac{\partial {\tilde\varepsilon}_f}{\partial \nu_f},
\label{eq:nuf2}\\
  - U_d & = & \frac{\partial{\cal E}}{\partial q_d}\cdot\frac{\partial q_d}
  {\partial {\nu_d}} + \frac{\partial{\cal E}}{\partial {\tilde\varepsilon}_f}
    \cdot\frac{\partial {\tilde\varepsilon}_f}{\partial \nu_d}.
\label{eq:nud}
\end{eqnarray}

First we derive analytic results from these equations in the weak
hybridization limit up to $O((V/W)^2)$ for arbitrary $U_f$ at special 
fillings:\ for $n_f=1$ and $n_d$ arbitrary; 
for $n_d=1$ and $n_f$ arbitrary; and finally for $n_f=n_d=1$. Similar results were 
obtained in Ref.\ [\onlinecite{Itai:variational}], but only for $U_f \rightarrow \infty$. 

We know that the interaction between conduction electrons suppresses 
charge fluctuations in the Hubbard subsystem. This influences the Kondo 
physics in the following ways: 

(i) $U_d$ shifts the Fermi energy of the conduction band. For $n_f=1$
and $n_d<1$ we get 
\begin{gather}
E_{\text{F}}(n_d, U_d) \approx  \left(\frac{n_d}{2} - \frac{1}{2}\right)q_dW  
\;\;\;\;\;\;\;\;\;\;\;\;\;\;\;\;\;\;\;\;\;\;\;\;\;\;\;\;\;\;\;\;\;\; \nonumber \\ 
\;\;\;\;\; +\; W\left[-\frac{1}{4} + \left(\frac{n_d}{2}-\frac{1}{2}\right)^2 
- 2\left(\frac{V}{W}\right)^2\frac{q_f}{q_d}\right]
\frac{\partial q_d}{\partial n_d}.
\label{eq:E_F} 
\end{gather}
The third term in the square brackets is the contribution of $d$-$f$ 
hybridization. Without it we recover the equation determining the 
Fermi energy of the Hubbard model for filling $n_d$. Note that the values 
of $\nu_f$ and $\nu_d$ in $q_f$ and $q_d$, respectively, should be 
taken from the solution of Eqs.\ (\ref{eq:nuf2}) and 
(\ref{eq:nud}). Equation (\ref{eq:E_F}) has no simple closed form 
solution for arbitrary $U_d$ except for the half-filled case, where 
$E_{\text{F}}(n_d=1,U_d) = U_d/2$. At other fillings 
we can expand $E_{\text{F}}(n_d, U_d)$ in the weak- or strong-coupling 
limit ($U_d\ll W$ or $U_d\gg W$) as
\begin{gather}
E_{\text{F}}(n_d, U_d) \approx E_{\text{F}}(n_d, 0) + \frac{n_d}{2} U_d + O(U_d^2/W)\label{eq:E_F-small-Ud}
\end{gather}
or
\begin{gather}
E_{\text{F}}(n_d, U_d) \approx E_{\text{F}}(n_d, 0) + \frac{n_d}{2} W + O(W^2/U_d),\label{eq:E_F-big-Ud}
\end{gather}
respectively. The shift of the Fermi energy is at most $W/6$ for $n=4/3$ 
(i.e., $n_d=1/3$), which is much smaller than the shift in the half-filled case. 

(ii) Switching on $U_d$ reduces the Kondo energy.\cite{Itai:variational} 
When $n_f = 1$, we can calculate $q_f$ and the total energy density for 
finite $U_d$ and for arbitrary $n_d$ (assuming $V \ll W$). Instead of Eqs.\ 
(\ref{q_f-symm}) and (\ref{eq:Esym}) we find 
\begin{equation}
q_f = \frac{n_d q_d}{4(V/W)^2}\; 
     \exp \left( -\frac{U_f}{16V^2/W}\right),\label{q_f-symm-Ud}
\end{equation}
and
\begin{gather}
   {\mathcal E} = \varepsilon_f + E_d(n_d, U_d) 
   - n_d q_d \frac{W}{2} \exp {\left\{-\frac{U_f}{16V^2/W}\right\}},
\label{eq:kondo-scale-Ud}
\end{gather}
where the second term of the right hand side is the energy of the decoupled 
correlated conduction band. Compared with Eqs.\ (\ref{q_f-symm}) and 
(\ref{eq:kondo-scale}), $q_f$ and the exponential Kondo scale 
are reduced by $q_d$, which is rather small when $n_d \approx 1$ (see below). 
For $n$ slightly less than the half-filled case this mechanism yields 
a significant mass enhancement. We get $q_d \sim 1/5$ for $n = 1.95$ and 
$U_d = 2.4 W$, which means that the effective mass
is five times bigger for these parameters than without $U_d$.

(iii) The most interesting effect of $U_d$ is the Mott transition which 
occurs in the Hubbard model at half filling ($n_d=1$). In the Gutzwiller-type 
treatment of $U_d$ it is known as the Brinkman-Rice transition. It occurs when 
$q_d$ becomes zero for a finite $U_d$. A similar transition may take place in 
the half-filled periodic Anderson-Hubbard model. In this model, however,
even if $n=2$, the Kondo physics may compete with Mott physics, 
$n_d$ and $n_f$ depend on $U_d$, $U_f$, $V$, and $\varepsilon_f$ owing to 
the $d$-$f$ hybridization,  and the conditions for the Mott transition may 
not be so simple as in the Hubbard model. In what follows we first show in 
the framework of the Gutzwiller treatment that the necessary conditions 
for the Mott transition is that both the $f$- and $d$-electron subsystem 
be half filled, i.e., $n_d=1$ and $n_f=1$ be fulfilled simultaneously, 
and moreover the system be in the Kondo regime. 

We see from Eq.\ (\ref{q_d}) that $q_d$ is zero only when $n_d=1$ 
and $\nu_d=0$. Similarly it follows from Eq.\ (\ref{q_f}) that $q_f$
vanishes only if $n_f=1$ and $\nu_f=0$. When $n_d=1$, the renormalization 
factor $q_d$ is simply $8\nu_d(1-2\nu_d)$, and Eq.\ (\ref{eq:nud}) gives 
\begin{gather}
\frac{U_d}{W} - \left[\frac{1}{4}+
2\left(\frac{V}{W}\right)^2\frac{q_f}{q_d}\right]8(1-4\nu_d)=0
\label{eq:nu_d-BR}
\end{gather}
for $V \ll W$ and $n_f$ arbitrary. The second term in the square brackets 
is the contribution of $d$-$f$ hybridization. Without it we recover the 
equation determining the optimum $\nu_d$ of the half-filled Hubbard model. 

It follows from this equation that $\nu_d$ goes to zero as $U_d$ approaches 
a finite critical value only if $q_f$ also approaches zero, and $q_f$ is of the 
same order as $q_d$. This situation can be realized only if $n_d=1$ and 
$n_f=1$ are simultaneously fulfilled, and moreover $q_f$ is given 
by Eq.\ (\ref{q_f-symm-Ud}), i.e., the system is in the Kondo regime.

When $n_d=1$ and $n_f=1$ are simultaneously satisfied, and the system 
is in the Kondo regime, the term in \eqref{eq:nu_d-BR} due to $d$-$f$ 
hybridization is independent of $\nu_d$ and is equal to $E_{\text{K}}/W$ 
[see Eqs.\ (\ref{eq:kondo-scale}),  (\ref{q_f-symm-Ud}), and 
(\ref{eq:kondo-scale-Ud})]. Equation (\ref{eq:nu_d-BR}) is easily solved to give
\begin{gather}
\nu_d = \frac{1}{4} - \frac{U_d}{8(W+4E_{\text{K}})},\label{nu_d-symm-Ud}
\end{gather}
which shows that $\nu_d$ decreases linearly as $U_d$ increases and reaches 
zero at $U_d^{\text{c}}=2(W+4E_{\text{K}})$. At this value of $U_d$, which --
owing to the coupling between the $d$- and $f$-electron subsystems -- is
slightly larger than the critical value in the Hubbard model 
($U_d^{\text{c}} = 2W$), the conduction band undergoes a Brinkman-Rice 
transition. Note that the exponentially small correction has been neglected 
in Ref.\ [\onlinecite{Itai:variational}]. Since $\nu_d = \nu_f = 0$ at 
this transition, all polarity fluctuations are suppressed, the effective $d$-$f$ 
hybridization ($\tilde{V} = V \sqrt{q_d q_f}$) as well as the Kondo energy 
scale become zero, that is, the Kondo effect is completely quenched. 
The system transforms from a Kondo insulator into a Mott insulator. 

Analytically we can claim only that the condition $n_d=n_f=1$ 
is realized in the symmetric point of the half-filled model, where 
$\varepsilon_f=-U_f/2+U_d/2$. Indeed, when $U_d$ is smaller than 
$U_d^{\text{c}}$ and is not very close to it, it is found
numerically that $n_d=n_f=1$ is realized only at the symmetric point, 
and thus one could expect that a Brinkman-Rice transition occurs only 
in the half-filled symmetric periodic Anderson-Hubbard model, and that 
the system becomes a Mott insulator for $U_d>U^{\text{c}}_d$ only if 
$\varepsilon_f=-U_f/2+U_d/2$. 

\begin{figure}[!ht]
\includegraphics[scale=0.5]{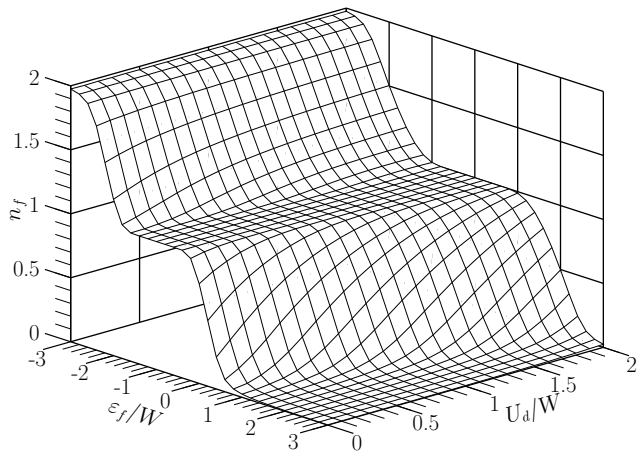}
\caption{\label{nf_Ud:fig}
The $f$-level occupancy vs.\ the $f$-level energy and 
$U_d$ at half filling for $V/W=0.1$ and $U_f/W=2$.}
\end{figure}

\begin{figure}[!ht]
\includegraphics[scale=0.5]{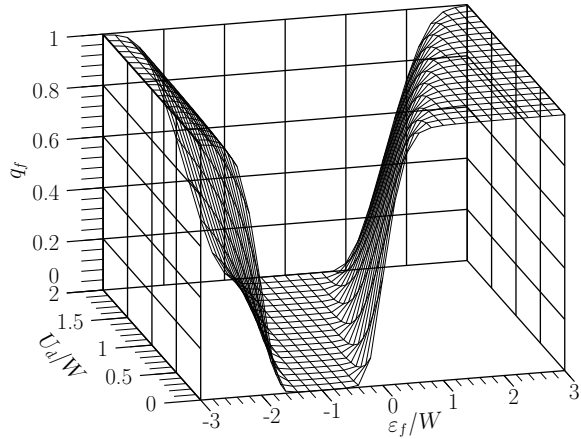}
\caption{\label{qf_Ud:fig}
The kinetic energy renormalization factor for $f$-electrons vs.\ 
$\varepsilon_f$ and $U_d$ at half filling for $V/W=0.1$ and $U_f/W=2$.}
\end{figure}

Contrary to this expectation we have found numerically that 
when $U_d$ is slightly smaller than the critical value, 
$n_f=n_d=1$ holds not only at the symmetric point, but -- within 
the limits of the numerical accuracy of our calculations, which was 
about $10^{-6}$ -- in a wide range of $\varepsilon_f$ within the 
$n_f \approx 1$ plateau. In order to find the extent of this
range, we display the $U_d$- and $\varepsilon_f$-dependence of the 
$f$-level occupancy and of the renormalization factor $q_f$ for 
$U_d \leq 2W$ at half filling in Figs.\ \ref{nf_Ud:fig} and 
\ref{qf_Ud:fig}, respectively.

It is clearly seen that the Kondo plateau, where $n_f \approx 1$ 
and $q_f \approx 0$, shifts towards higher energies owing to 
the shift of the Fermi energy by $U_d/2$, and the center of the 
plateau is located indeed at $\varepsilon_f=-U_f/2+U_d/2$ as expected 
from  Eq.\ (\ref{eq:E_F}). The condition for the Kondo regime can be 
written similarly to Eq.\ (\ref{eq:plateau}) as
\begin{equation}
-U_f+E_{\text{F}}(n_d, U_d)+a\Delta_f \lesssim 
\varepsilon_f \lesssim E_{\text{F}}(n_d,U_d)-a\Delta_f.
\label{eq:plateau-Ud}
\end{equation}
Note that the center of the plateau is at the 
center of the noninteracting $d$-band, when $U_d=U_f$. 
In other words, the $f$-level does not need to lie low enough compared to 
the conduction band to show heavy-fermion behavior.

Another remarkable feature is that the plateau widens as $U_d$ approaches 
the critical value $U_d^{\text{c}}$. At $U_d = 2W$, it is situated in the 
range $-U_f+ U_d/2 \lesssim \varepsilon_f \lesssim U_d/2$. That means
that the narrowing of the plateau compared to the impurity model given 
by $a\Delta_f$ in Eq.\ (\ref{eq:plateau}) gets 
remarkably smaller close to $U_d^{\text{c}}$. This is probably 
due to the formation of the Hubbard subbands and the drastic variation 
of the density of states at the Fermi energy near the transition point.

Numerical calculations give $n_d=n_f=1$
at $U_d^{\text{c}}$ on the whole Kondo plateau.
This indicates that -- at least within the Gutzwiller-type treatment 
of correlations -- both $n_d$ and $n_f$ are fixed to exactly unity in 
the half-filled model as we approach $U_d^{\text{c}}$ and 
the condition for Kondo behavior is satisfied. The renormalization factors,
$q_d$ and $q_f$ vanish simultaneously, the $d$-$f$ hybridization 
is completely suppressed, so is the Kondo effect, and a Mott 
transition takes place. This transition in the conduction electron 
subsystem is robust, it is the dominant feature of the half-filled 
model. 

Our finding that the Brinkman-Rice transition and the Mott
insulating state are not restricted to the symmetric model is corroborated 
by calculations for $U_d>U^{\text{c}}_d$. The numerical variational 
calculation yields meaningless negative values for $\nu_d$ in the whole 
interval $-U_f + U_d/2 \lesssim \varepsilon_f \lesssim U_d/2$. 
Note that for $\varepsilon_f$ outside this interval we can carry out 
the numerical calculations for arbitrary large $U_d$ without any difficulty. 

Next we show that the $d$-$f$ hybridization prevents the Mott transition 
when $n<2$ (or for $n > 2$). In this case the term coming from the $d$-$f$ hybridization 
in Eq.\ (\ref{eq:nu_d-BR}) becomes large, if  $q_d \rightarrow 0$, since 
$q_f$ is always finite for $n_f<1$, and thus there exists no such solution 
for $\nu_d$ (or $q_d$), which approaches zero at a finite $U_d$. 
Charge fluctuations on the $d$-orbitals are thus not completely suppressed.
A finite $\nu_d$ indicates the existence of a Fermi surface, since 
$q_d$ is identified with the discontinuity at the Fermi wavenumber in the 
single-particle occupation number.\cite{Gutzwiller:original} 
This can be understood as follows:\ even if the correlated 
conduction band is half filled and $U_d$ is large enough, so that the 
conduction band is separated into Hubbard subbands and  
the Fermi level lies within the $f$-band located in the Hubbard gap, 
the $d$-electrons are taking part in the formation of the Fermi 
surface via $d$-$f$ hybridization. 

\begin{figure}[!htb]
\includegraphics[scale=0.5]{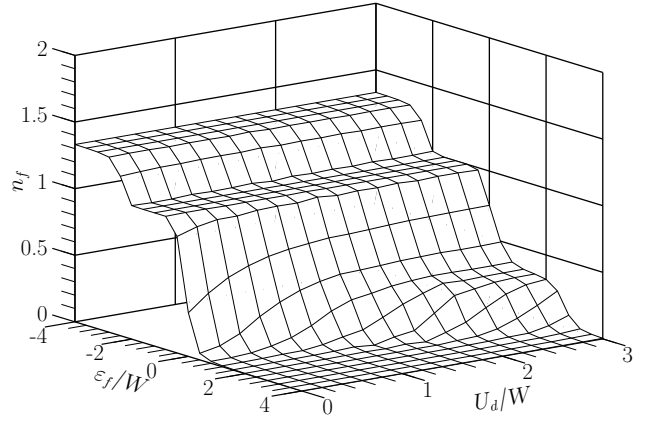}
\caption{\label{nf_Ud-8/6:fig}
The $f$-level occupancy vs.\ $f$-level energy and 
$U_d$ at 1/3 filling ($n=4/3$), for $V/W=0.1$ and $U_f/W=2$.}
\end{figure}

The results of the numerical calculations in the 1/3-filled case 
($n=4/3$) are shown for $0 \leq U_d \leq 3W$ in Fig.\ \ref{nf_Ud-8/6:fig}. 
We observe that one more plateau appears at higher $f$-level energies, 
above the bare conduction band, when $U_d \gtrsim 2W$, corresponding 
to $n_d \approx 1$. Its formation indicates 
that two separate Hubbard subbands are formed above this critical value 
of $U_d$. The plateau appears when $\varepsilon_f$ is located between 
the two subbands, i.e., in the Hubbard gap. The Fermi level is located in 
the $f$-band in this situation. The center of the plateau is approximately 
at $U_d/2$, which indicates that the upper and lower subbands are centered 
at 0 and $U_d$, respectively, i.e., the location of the subbands is the 
same as in the Hubbard model.

We observe furthermore that the $n_f \approx 1$ plateau hardly shifts as 
$U_d$ increases. Since according to Eq.\ (\ref{eq:E_F-big-Ud}) the Fermi 
energy only weakly depends on $U_d$ away from half filling, we find again
the condition given in \eqref{eq:plateau-Ud} for the Kondo regime.

It is worth mentioning here what happens when the 
system is more than half filled. The answer can be obtained without any further 
calculation from electron-hole symmetry. 
A model with $n$ electrons can be mapped onto a model with 
$n$ holes ($4-n$ electrons) by the transformation: 
\begin{gather}
\hat d_{j\sigma}^{\dagger} \rightarrow 
e^{i\varphi_{j}} \hat d_{{\text{h}}j\bar\sigma}, \; \; \; \; 
\hat d_{j\sigma} \rightarrow 
e^{-i\varphi_{j}} \hat d^{\dagger}_{{\text{h}}j\bar\sigma},\nonumber\\
\hat f_{j\sigma}^{\dagger} \rightarrow 
 - e^{i\varphi_{j} }\hat f_{{\text{h}}j\bar\sigma}, \; \; \; \;
\hat f_{j\sigma} \rightarrow 
- e^{-i\varphi_{j} }\hat f^{\dagger}_{{\text{h}}j\bar\sigma},\label{eq:ph-tr}
\end{gather}
where the index (h) refers to holes, 
and $\bar\sigma=-\sigma$. If the kinetic energy of conduction electrons
is written in Wannier representation,
\begin{equation}
       \sum_{\boldsymbol{k},\sigma}\varepsilon_d(\boldsymbol{k})
       \hat{d}_{\boldsymbol{k}\sigma}^{\dagger}
	   \hat{d}^{\phantom \dagger}_{\boldsymbol{k}\sigma} =
	    \sum_{ij\sigma}t_{ij}  \hat{d}_{i\sigma}^{\dagger}
	   \hat{d}^{\phantom \dagger}_{j\sigma} ,
\end{equation}
and the phase factor is chosen in the form 
$\varphi_{j}=\boldsymbol Q\cdot{\boldsymbol r}_{j}$,
it is easily seen that the kinetic energy term is transformed
into 
\begin{equation}
       \sum_{\boldsymbol{k},\sigma}\varepsilon_{\text{h}d}(\boldsymbol{k})
       \hat{d}_{{\text{h}}\boldsymbol{k}\sigma}^{\dagger}
	   \hat{d}^{\phantom \dagger}_{{\text{h}}\boldsymbol{k}\sigma},
\end{equation}
where
\begin{equation}
     \varepsilon_{\text{h}d}(\boldsymbol{k}) = - \varepsilon_{d}(\boldsymbol{k} + {\boldsymbol{Q}}).
\end{equation}
Assuming that $t_{ii} = 0$, the center of the band sets the zero of energy.
The term describing hybridization is invariant under this transformation, while the
on-site energy of $f$-levels and the on-site interaction terms give rise to energy 
shifts. Therefore the Hamiltonian written in terms of the creation and annihilation 
operators of holes has the same form as for electrons with shifted energies for the
$d$- and $f$-electrons, and an overall energy shift:
\begin{gather}
{\mathcal H}_{\text{e}}\left(\varepsilon_d(\boldsymbol k),
\varepsilon_f,V,U_d,U_d\right)\longrightarrow\;\;\;\;\;\;\;\;\;\;\;\nonumber\\
{\mathcal H}_{\text{h}}\left(\varepsilon_{\text{h}d}({\boldsymbol k})-U_d,\;
\varepsilon_{\text{h}f},V,U_d,U_f\right)+ E_0,\label{eq:ph-tr-hamiltonian}
\end{gather}
where $\varepsilon_{\text{h}f}= -\varepsilon_f - U_f$ 
and $E_0=(2\varepsilon_f + U_f + U_d)N$. If the energy levels are measured
from $-U_d$ the Hamiltonian in hole representation becomes
\begin{equation}
{\mathcal H}_{\text{h}}\left(\varepsilon_{\text{h}d}({\boldsymbol k}),\;
\bar{\varepsilon}_{\text{h}f},V,U_d,U_f\right)+ \bar{E}_0,\label{eq:ph-tr-hamiltonian-2}
\end{equation}
with $\bar{\varepsilon}_{\text{h}f}= -\varepsilon_f - U_f + U_d$ and 
\begin{equation}
\bar{E}_0=-(2\varepsilon_d + U_d)N_{\text{h}} + (2\varepsilon_f + U_f + 2\varepsilon_d + U_d)N,
\end{equation}
where $N_{\text{h}}$ is the total number of holes. Provided that 
$\varepsilon_{\text{h}d}({\boldsymbol k}) \equiv - \varepsilon_d({\boldsymbol k}+ \boldsymbol{Q})=\varepsilon_d({\boldsymbol k})$ 
for a certain ${\boldsymbol Q}$, as is the case for the one-dimensional model
with nearest-neighbor hopping, or when a constant density of states is assumed, 
then the dispersion curve of $d$-holes is the same as for $d$-electrons and
the results obtained in the electron representation can be applied to holes
when the energy shifts are taken into account. 

Using this transformation, the results for $n > 2$ can be obtained straightforwardly from 
those for $n_{\text{h}} = 4-n < 2$. We can get, for example, the Fermi energy of the correlated 
conduction band for $n_d = n-1>1$ ($n_f=1$) from that for $n_{\text{h}d} = 2-n_d < 1$ 
[see Eq.\ (\ref{eq:E_F})] by first shifting the origin of the energy by $-U_d$, and then 
reversing the energy axis. We get  
\begin{gather}
E_{\text F}(n_d,U_d)= - \big[E_{\text F}(n_{\text{h}d},U_d) - U_d\big],\label{eq:E_F-n>2}
\end{gather}
from which for $n_d>1$
\begin{gather}
E_{\text F}(n_d,U_d=0)= - E_{\text F}(2 - n_d,U_d=0).
\end{gather} 
The equation giving the shift of the Fermi energy for $n_d > 1$ is thus
\begin{gather}
E_{\text F}(n_d,U_d) \approx E_{\text F}(n_d,0) - \frac{2-n_d}{2}W + U_d  + O(W^2/U_d)
\end{gather}
instead of Eq.\ (\ref{eq:E_F-big-Ud}). This shows that the shift of the Fermi 
energy owing to $U_d$ for $n_d>1$ is larger than that at half-filling.

The condition on $\varepsilon_f$ for the Kondo regime is obtained for $n>2$ 
as follows: The condition on the $f$-hole level $\bar\varepsilon_{\text{h}f}$ 
for $n_{\text{h}}<2$ is formally the same as for electrons 
[see Eq.\ (\ref{eq:plateau-Ud})], since the Hamiltonian has the same form, i.e., 
\begin{equation}
-U_f+E_{\text{F}}(n_{\text{h}d}, U_d)+a\Delta_f \lesssim
\bar\varepsilon_{\text{h}f} \lesssim E_{\text{F}}(n_{\text{h}d},U_d)-a\Delta_f.
\end{equation}
The condition on the $f$-electron level for the Kondo regime for $n>2$ is simply 
obtained by rewriting this condition for the original $\varepsilon_f$ using 
$\bar\varepsilon_{\text{h}f}=-\varepsilon_f-U_f+U_d$. We get
\begin{gather}
-U_f+ U_d - E_{\text{F}}(n_{\text{h}d}, U_d)+a\Delta_f\lesssim  \;\;\;\;\;\;\;\;\;\;\;\;
\;\;\;\;\;\;\;\;\ 
\nonumber\\
\;\;\;\;\;\;\;\;\;\;\varepsilon_f\lesssim U_d - E_{\text{F}}(n_{\text{h}d},U_d)-a\Delta_f.
\end{gather}
Since according to Eq.\ (\ref{eq:E_F-n>2}) $U_d - E_{\text{F}}(n_{\text{h}d}, U_d)$ ($n_{\text{h}d}<1$) is the Fermi energy of the interacting conduction band, 
$E_{\text{F}}(n_d, U_d)$, for $n_d=2-n_{\text{h}d}>1$, the condition takes the 
same form given in Eq.\ (\ref{eq:plateau-Ud}) for all fillings.
For $n>2$, the shift of the $n_f \approx 1$ plateau is thus even larger than in 
the half-filled case, it may appear above the bare conduction band.

\subsection{Comparison with the results by exact diagonalization}

Now, we  compare the results of exact diagonalization with those obtained by
variational calculation. Note that we will discuss only the case $n>1$, i.e., 
more than six electrons on a six-site chain. The quarter-filled case is not 
interesting from the point of view of Kondo physics, because the conduction 
band is exhausted when $n_f=1$.

\begin{figure}[!t]
\includegraphics[scale=0.5]{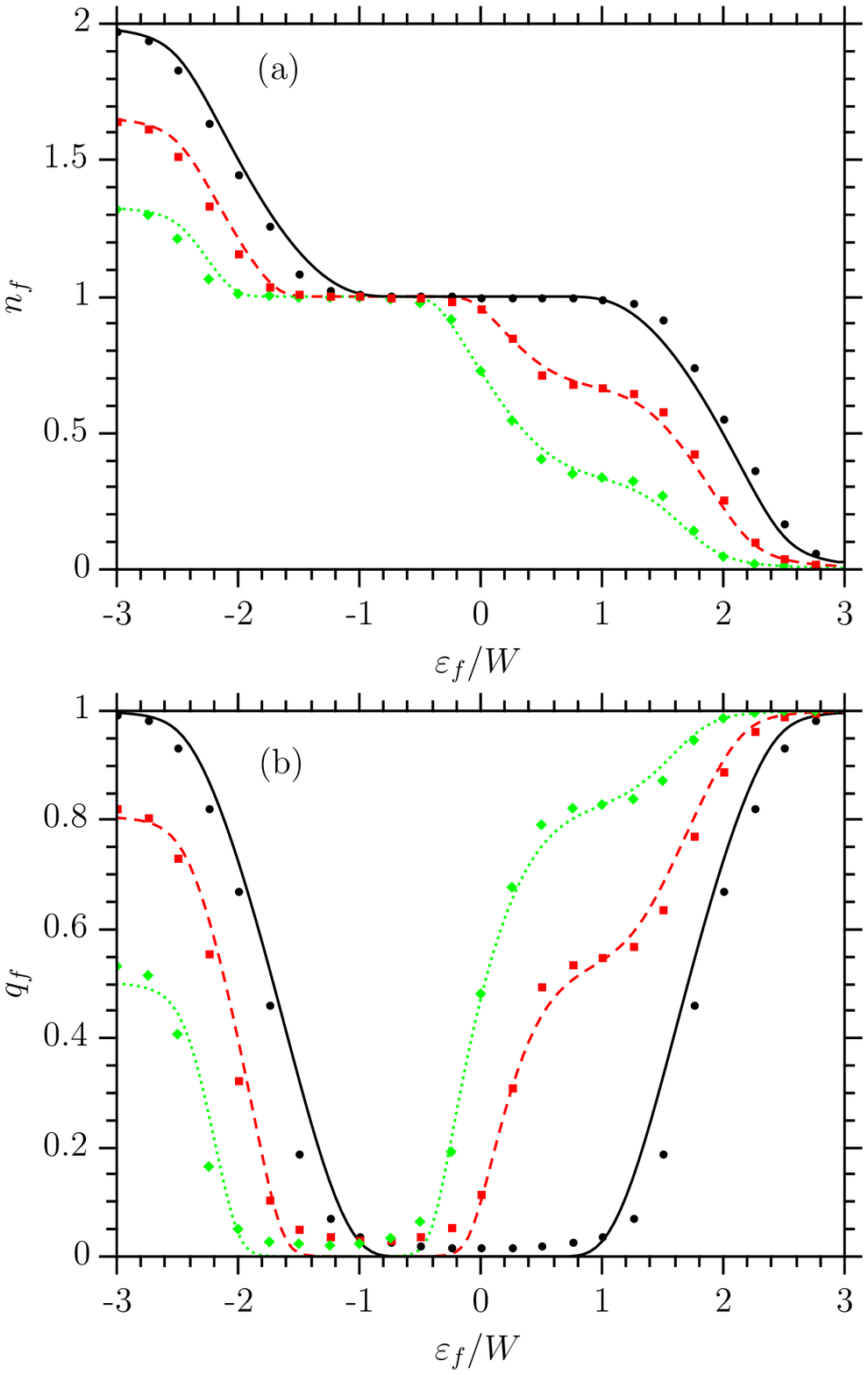}
\caption{\label{filling2:fig} (Color online) (a) The $f$-level occupancy vs. 
$\varepsilon_f$ at $U_f/W=U_d/W=2$ and  $V/W = 0.1$. The curves 
are obtained using the Gutzwiller method, while the symbols are the results of 
exact diagonalization. The electron number per site is 2 (half filling) 
(black continuous line, {\tiny\textbullet}), 5/3 (red dashed line, 
{\tiny{$\blacksquare$}}), and 4/3 (green dotted line, {\tiny{$\blacklozenge$}}), 
respectively. (b) The renormalization factor $q_f$. The notation is 
the same as in panel (a).}
\end{figure}

The values of $n_f$ obtained by both methods are shown for several 
fillings at $U_f = U_d = 2W$ in Fig.\ \ref{filling2:fig}(a). The overall 
agreement between the results of the two methods demonstrated earlier 
remains good for finite $U_d$. It is remarkable that the agreement is 
even better than for $U_d=0$.

The shift of the  $n_f \approx 1$ plateau due to $U_d$ is observed in both 
methods, showing that the shift of the plateau is not 
an artefact of the Gutzwiller approximation, and may be observable in some 
materials, where the conduction electrons are strongly correlated, i.e., 
they may exhibit heavy-fermion behavior despite the fact that the 
bare $f$-level does not lie below the conduction band. 

The formation of two separate plateaus corresponding to $n_f \approx 1$ 
and  $n_d \approx 1$ is also observed in both methods. The formation of 
the Hubbard subbands owing to $U_d$ in the conduction-electron subsystem is 
thus confirmed by exact diagonalization, too. We carried 
out the comparison also at $U_f = U_d =5 W$ for $n=5/3$, and found 
that the agreement between the two methods is almost perfect for $n_f$.

\begin{figure}[!t]
\includegraphics[scale=0.5]{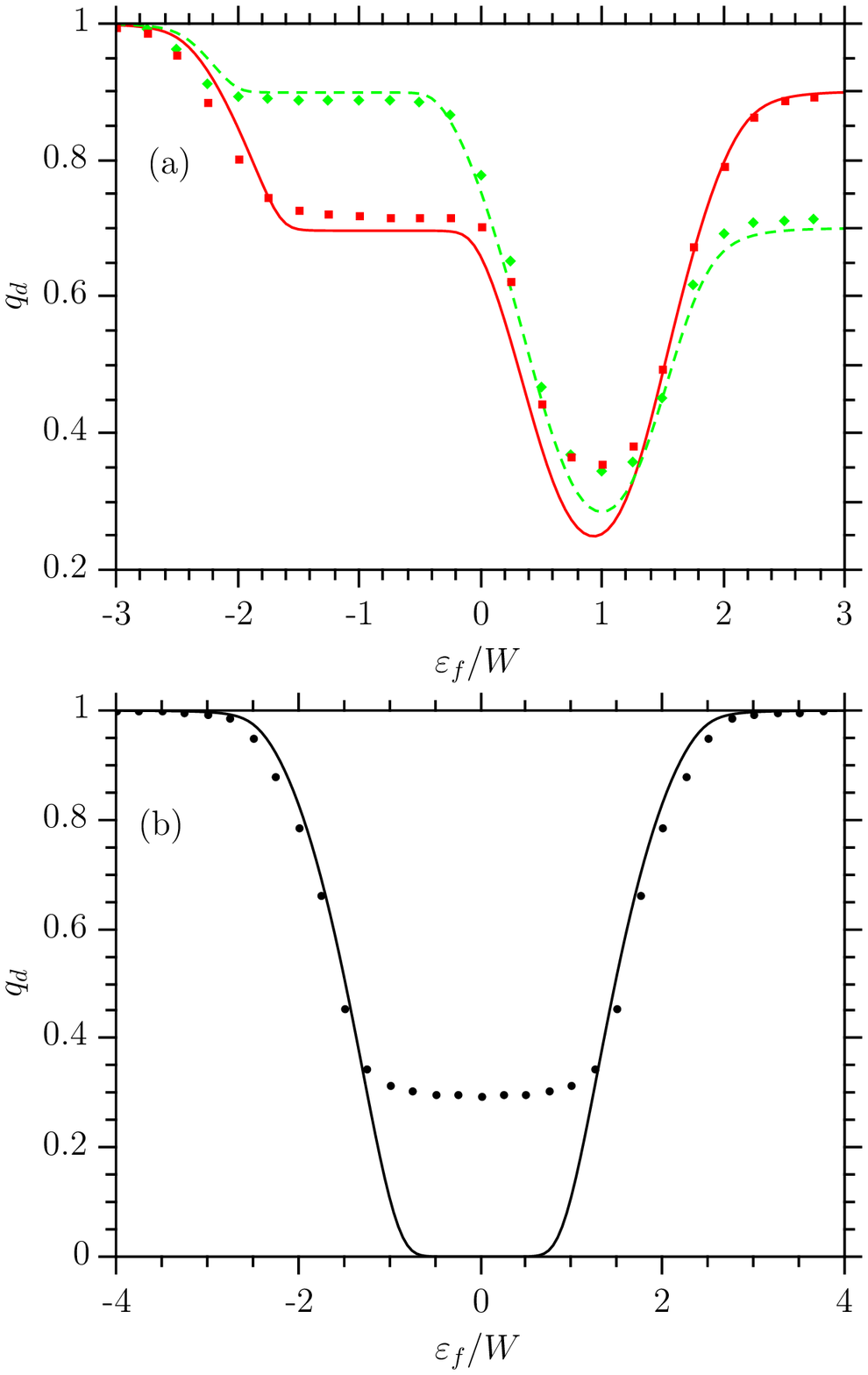}
\caption{\label{qd:fig} (Color online) The kinetic energy renormalization 
factors of the conduction electrons vs.\ $\varepsilon_f$ at 
$U_f/W=U_d/W=2$.  The curves are obtained by the Gutzwiller 
method, the symbols show the values calculated with exact diagonalization 
using Eq.\ (\ref{q_d}). (a) The results for $n=5/3$ (the red dashed line, 
{\tiny{$\blacksquare$}}) and $n$=4/3 (the green dotted line,
{\tiny{$\blacklozenge$}}). The hybridization is $V/W = 0.1$ in all cases. 
(b) The results for half filling, $n=2$ (the black continuous line, 
{\tiny\textbullet}).} 
\end{figure}

Figure \ref{qd:fig} shows the kinetic energy renormalization factor 
of conduction electrons ($q_d$) for three different fillings. The 
agreement between the two methods are fairly good in panel (a), while 
a marked difference is seen in panel (b), i.e., at half filling. At this 
point we should note that the two methods are complementary. The Gutzwiller 
method is exact for large dimensions, while the exact diagonalization was 
performed on chains. We argue that the difference between the results 
obtained by the two methods at half filling is due to the unusual 
behavior of the one-dimensional half-filled Hubbard model. 

The one-dimensional Hubbard model can be solved exactly by Bethe 
ansatz.\cite{Lieb-Wu} At half filling its ground state is conducting only 
for $U_d = 0$ and insulating for any nonzero $U_d$. On the other hand, 
in higher dimensions the half-filled Hubbard model is expected to remain 
metallic until a finite critical $U_d^{\text{c}}$, where the Mott 
transition takes place. Therefore, when we compare the results 
obtained by the Gutzwiller method and by exact diagonalization of the
Hamiltonian of a chain displayed in Fig.\ \ref{qd:fig}(b) and interpret 
the difference, we have to keep in mind the fundamental difference between the physics of one- 
and higher dimensional half-filled Hubbard models.

The Gutzwiller method gives not only a vanishing valence fluctuation 
on $f$-orbitals, $\nu_f=0$, and consequently $q_f=0$ 
at $U_d = 2W + 8E_{\text{K}}$, but the same is true for the 
conduction-electron subsystem: also $\nu_d$ and $q_d$ vanish at the
Brinkman-Rice transition. On the other hand, exact diagonalization 
gives a finite $q_d$ in agreement with the known behavior of the 
one-dimensional half-filled Hubbard model, where $\nu_d$ is 
finite for arbitrary $U_d$.\cite{Woynarovich} We believe that the 
disagreement between the predictions of the Gutzwiller method
and of exact diagonalization seen in Fig.\ \ref{qd:fig}(b) 
is thus the consequence of the different behavior of the one-dimensional
and higher dimensional periodic Anderson-Hubbard models.

\section{Conclusions}

In this paper we considered an extended periodic Anderson modell, the so-called
periodic Anderson-Hubbard model with on-site Coulomb repulsion in the 
conduction-electron subsystem. Our main aim was to investigate how the additional 
repulsive interaction between conduction electrons influences the Kondo
regime, and how the Kondo physics and Mott physics compete. For this study we 
calculated the average number of $f$- and $d$-electrons per site, $n_f$ and
$n_d$, and the probability of double occupancy in both subsystems, $\nu_f$ and
$\nu_d$, using the Gutzwiller variational method. In order to check the 
reliability of this method, we also performed exact diagonalization on 
relatively short chains. Since to our best knowledge no such comparison was 
presented even for the original periodic Anderson model, we also present
results for the original periodic Anderson model. 

A rather good agreement was found between the results of the 
two methods in the original model as far as the location of the Kondo 
and valence-fluctuation regimes are concerned. A subtle difference was, 
however, found near the boundary of the Kondo plateau. Namely, while the 
results of the Gutzwiller method exhibit an exponential dependence of 
the double occupancy on the characteristic combination of the couplings, 
$V^2/U_f$, those of exact diagonalization show a power-law behavior. 
This will be the subject of subsequent studies. 

The situation is somewhat different for the extended model. Both methods
indicate that when the on-site Coulomb repulsion between conduction electrons ($U_d$) 
is switched on, in the half-filled case the heavy-fermion regime shifts towards 
higher energies of the bare $f$-level by $U_d/2$ in accordance with the shift 
of the Fermi energy owing to $U_d$. A marked 
difference appears, however, between the results provided by the two methods, when 
$U_d$ is of the order of $2W$. The Gutzwiller method indicates that 
both $n_d$ and $n_f$ are fixed to unity for a wide range of the $f$-level
energies in the half-filled model at a critical value of $U_d$, and a
robust Brinkman-Rice transition takes place to a Mott insulator. Although 
the Kondo effect is enhanced, when the Anderson-Hubbard system approaches 
the critical point, this effect is completely suppressed right at the transition,
and all charge fluctuations are suppressed. Even though the exact diagonalization
on chains does not reproduce this result, we believe that the Gutzwiller 
method describes correctly the scenario in higher dimensional systems, and 
the different behavior found for linear chains is simply a consequence
of the anomalies of low-dimensional systems.

When the electron system is less than half filled, the Mott transition is 
suppressed by the $d$-$f$ hybridization, and besides the Kondo plateau 
($n_f \approx 1$) another plateau appears at $n_d \approx 1$, provided that
$U_d \gtrsim 2W$, i.e., when the conduction band is split into a lower and
upper Hubbard band. The results provided by the two methods are in 
surprisingly good agreement in this case, in particular when correlations 
are strong. The shift of the heavy-fermion regime towards higher
bare $f$-level energies owing to $U_d$ is small compared to that in 
the half-filled case, because the shift of the Fermi energy due to 
$U_d$ is at most $n_dW/2$ for $n_d<1$. On the other hand, 
when the electron system is more than half filled, the shift of the 
Kondo regime with $U_d$ is much larger, since the shift of the Fermi energy 
is also larger than that  in the half-filled case.

\acknowledgments{This work was supported in part by the
Hungarian Research Fund (OTKA) through Grant No.~T 68340.} We acknowledge 
F. Woynarovich for fruitful discussions.

\end{document}